\documentclass[fp,twocolumn]{jpsj3}

\title{Surface Structure and Surface State
  of a Tight-Binding Model\\
  on a Diamond Lattice}

\author{Katsunori Kubo}
\inst{Advanced Science Research Center,
  Japan Atomic Energy Agency, Tokai, Ibaraki 319-1195, Japan}

\date{\today}

\abst{
Surface states of a tight-binding model with nearest-neighbor hopping
on a diamond lattice of finite thickness are investigated.
We consider systems with (001), (110), and (111) surfaces.
Even if the surface direction is fixed,
there is freedom to choose the surface structure
due to the two-sublattice nature of a diamond lattice.
The existence of surface states is governed
by the topology of the matrix elements of the bulk Hamiltonian.
In this sense, the existence of surface states is determined
by the bulk Hamiltonian.
However, the matrix elements depend on the choice of the unit cell,
which should be chosen to conform to the surface structure.
Thus, the surface states depend on the surface structure.
We find that for each surface direction,
there are two choices of surface structure,
and depending on the chosen structure,
the corresponding surface states emerge
in distinct regions of the surface Brillouin zone.
}

\begin{document}
\maketitle

\section{Introduction}
Topological semimetals have garnered significant attention
due to their unique and intriguing properties.
They are characterized by topologically protected degeneracies
in the electronic band structure.
These degeneracies can occur at points,
called Dirac or Weyl points~\cite{Wallace1947, Novoselov2005, Murakami2007, Wan2011, Kubo2024JPSJ, Kubo2024PRB},
or along continuous lines, called nodal
lines~\cite{Chadi1975, Heikkila2011, Burkov2011, Weng2015, Kubo2025}.

To realize topological semimetals,
at least two electronic bands are necessary,
as they are defined by the degeneracy of the bands.
For instance,
in the presence of Rashba spin-orbit coupling~\cite{Bychkov1984},
the spin degeneracy in the electronic bands is lifted,
resulting in two distinct bands.
In this scenario, spin degeneracy is preserved
at specific $\mib{k}$ points in the presence of the time-reversal symmetry.
These points are Weyl points,
and edge states originating from them have been discussed~\cite{Kubo2024JPSJ}.
In a multiorbital system, such multiple bands naturally arise.
As a typical example,
an $e_g$ orbital model has been studied,
and octupolar edge (surface) states originating
from Dirac points (nodal lines) were reported
for a square lattice~\cite{Kubo2024PRB}
(for a simple cubic lattice~\cite{Kubo2025}).

\begin{figure}
  \begin{center}
    \includegraphics[width=0.7\linewidth]
      {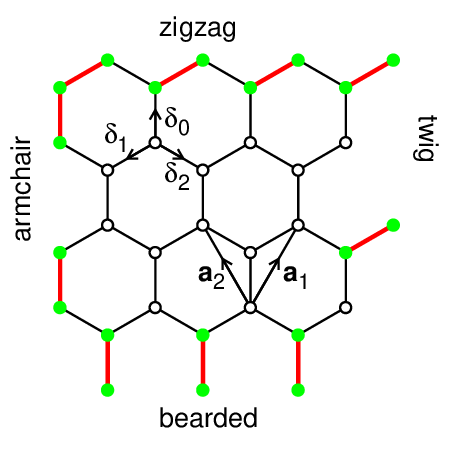}%
  \end{center}
  \caption{
    (Color online)
    Edge structures (armchair, twig, bearded, and zigzag)
    of a honeycomb lattice.
    Each bold red line connects two sites (solid green circles)
    constituting a unit cell on an edge.
    The primitive translation vectors are
    $\mib{a}_1=a(\sqrt{3},3)/2$ and $\mib{a}_2=a(-\sqrt{3},3)/2$,
    where $a$ is the bond length.
    The vectors connecting nearest-neighbor sites are
    $\mib{\delta}_0=a(0,1)$, $\mib{\delta}_1=a(-\sqrt{3},-1)/2$,
    and $\mib{\delta}_2=a(\sqrt{3},-1)/2$.
    \label{graphene_edges}}
\end{figure}

In the examples above,
the onsite degrees of freedom are used to realize two bands.
However, sublattice degrees of freedom can be employed
to realize multiple bands.
A prototypical example is the tight-binding model
with nearest-neighbor hopping on a honeycomb lattice,
which serves as a model for graphene.
This honeycomb lattice model possesses Dirac points~\cite{Wallace1947},
and edge states have been studied~\cite{Klein1994,Fujita1996, Ryu2002, Wakabayashi2010, Xia2023}.
Due to the sublattice degrees of freedom,
the edge shape is not uniquely determined,
even when the edge direction is fixed (see Fig.~\ref{graphene_edges}).
Consequently, the edge state depends not only on the edge direction
but also on the edge shape.
These edge states have been experimentally observed
in graphene~\cite{Kobayashi2005, Kobayashi2006}
and photonic crystals~\cite{Plotnik2014, Xia2023}.
In the next section,
we will review the edge states of the honeycomb lattice model.

In this study,
we extend the above findings of the honeycomb lattice model
to a three-dimensional system, namely, a diamond lattice.
The tight-binding model with nearest-neighbor hopping on a diamond lattice
possesses nodal lines~\cite{Chadi1975}.
In the presence of such nodal lines,
states localized well around the surfaces are expected in a lattice
with finite thickness~\cite{Heikkila2011, Burkov2011, Weng2015}.
Indeed, such surface states have been reported
for (111) surface~\cite{Takagi2008, Takahashi2013, Hirashima2016} and
for (001) surface~\cite{Hirashima2016}.
However, the potential variety of surface structure
has not been thoroughly explored.
Here, we investigate the surface states
for all the (001), (110), and (111) surfaces,
paying particular attention to the surface structures
arising from the two-sublattice nature of the diamond lattice.
Extensions of the edge states in the honeycomb lattice model
to higher dimensions are also discussed in Ref.~\citen{Takagi2009}.

\section{Honeycomb lattice}
In this section, we review the edge states
of the honeycomb lattice model.
The discussion will provide a useful foundation
for understanding the surface states of the diamond lattice model
in the next section.
Specifically, the formulations presented here can be readily extended
to the diamond lattice model.
The honeycomb lattice is illustrated in Fig.~\ref{graphene_edges}.
The edge states of the honeycomb lattice model were first discussed
for the bearded edge~\cite{Klein1994}.
Subsequently, the armchair and zigzag edges were studied
in Ref.~\citen{Fujita1996}.
A topological explanation for the existence of these edge states
was later provided in Ref.~\citen{Ryu2002}.
The twig edge was proposed thereafter in Ref.~\citen{Wakabayashi2010}.
We present a unified formulation for these four edge configurations.
Different edge shapes are realized
by selecting the appropriate edge direction and unit cell.

The primitive translation vectors are
$\mib{a}_1=a(\sqrt{3},3)/2$ and $\mib{a}_2=a(-\sqrt{3},3)/2$,
where $a$ is the bond length.
The vectors connecting nearest-neighbor sites are
$\mib{\delta}_0=a(0,1)$, $\mib{\delta}_1=a(-\sqrt{3},-1)/2$,
and $\mib{\delta}_2=a(\sqrt{3},-1)/2$.
Note that alternative choices for the primitive translation vectors
are possible.
For example, one could choose
$\mib{a}_1$ and $\mib{a}_1-\mib{a}_2=a(\sqrt{3},0)$
as primitive translation vectors.
Here, we define $\mib{a}^{(\mu)}_i=\mib{\delta}_{\mu}-\mib{\delta}_i$.
Then, $\{ \mib{a}^{(\mu)}_0, \mib{a}^{(\mu)}_1, \mib{a}^{(\mu)}_2 \}$
is the set of the zero vector $\mib{a}_0 \equiv (0,0)$
and the primitive translation vectors.
In particular, 
$\{ \mib{a}^{(0)}_0, \mib{a}^{(0)}_1, \mib{a}^{(0)}_2 \}
=\{ \mib{a}_0, \mib{a}_1, \mib{a}_2 \}$.

\begin{figure}
  \begin{center}
    \includegraphics[width=0.8\linewidth]
      {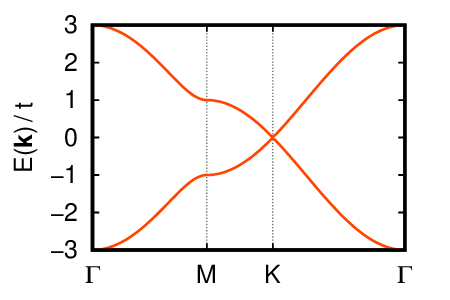}%
  \end{center}
  \caption{
    (Color online)
    Energy dispersion of the honeycomb lattice model.
    The high symmetry points are
    $\Gamma=(0,0)$, $M=2\pi(0,1/3)/a$, and $K=2\pi(1/3\sqrt{3},1/3)/a$.
    \label{graphene_dispersion}}
\end{figure}

A nearest-neighbor site of the site $\mib{r}$ on the $A$ sublattice
is $\mib{r}+\mib{\delta}_i=\mib{r}-\mib{a}^{(\mu)}_i+\mib{\delta}_{\mu}$,
which lies on the $B$ sublattice.
Then, if we choose unit cells
composed of $\{ \mib{r}, \mib{r}+\mib{\delta}_{\mu} \}$,
the Hamiltonian, in the absence of spin degrees of freedom,
is written as
\begin{equation}
  \begin{split}
    H
    &=
    t\sum_{\mib{r} i} c_{\mib{r} A}^{\dagger}c_{\mib{r}-\mib{a}^{(\mu)}_i B}+\text{h.c.}\\
    &=
    t \sum_{\mib{k} i}
    e^{-i\mib{k}\cdot \mib{a}^{(\mu)}_i}
    c_{\mib{k} A}^{\dagger}c_{\mib{k} B}+\text{h.c.}\\
    &=
    \sum_{\mib{k}}
    c_{\mib{k}}^{\dagger}
    [h_x^{(\mu)}(\mib{k})\sigma_x+h_y^{(\mu)}(\mib{k})\sigma_y]
    c_{\mib{k}}\\
    &=
    \sum_{\mib{k}}
    c_{\mib{k}}^{\dagger}
    H^{(\mu)}(\mib{k})
    c_{\mib{k}},
  \end{split}
  \label{eq:Hamiltonian}
\end{equation}
where
$t$ is the hopping integral,
$c_{\mib{r} \sigma}$ is the annihilation operator
for the electron on sublattice $\sigma$ in the unit cell at $\mib{r}$,
$c_{\mib{k} \sigma}$ is the Fourier transform of it,
$c_{\mib{k}}=(c_{\mib{k} A}, c_{\mib{k} B})^{\text{T}}$,
h.c. stands for the Hermitian conjugate terms,
$\sigma_{\alpha}$ denotes the $\alpha$ component of the Pauli matrix,
and
\begin{align}
  h^{(\mu)}_x(\mib{k}) &= t\sum_{i}\cos\mib{k}\cdot\mib{a}^{(\mu)}_i,
  \label{eq:hx}\\
  h^{(\mu)}_y(\mib{k}) &= t\sum_{i}\sin\mib{k}\cdot\mib{a}^{(\mu)}_i.
  \label{eq:hy}
\end{align}
Note that the model has a chiral symmetry:
$\{ H^{(\mu)}(\mib{k}), \sigma_z  \} = H^{(\mu)}(\mib{k}) \sigma_z+\sigma_z H^{(\mu)}(\mib{k}) = 0$
and $\sigma_z \sigma_z^{\dagger} = I$ with $I$ being the unit matrix.
In the following, we represent the choice of the unit cells by the index $\mu$.
The Hamiltonian expressed using a different choice of unit cells
is related to the original one by ``the gauge transformation'':
$\sum_{i} e^{-i\mib{k}\cdot \mib{a}^{(\nu)}_i}
= e^{i\mib{k} \cdot \mib{a}^{(\mu)}_{\nu}}\sum_i e^{-i\mib{k}\cdot \mib{a}^{(\mu)}_i}$.
Thus, the energy dispersion of the bulk Hamiltonian does not depend
on the choice of unit cells
and is shown in Fig.~\ref{graphene_dispersion}.
The bands degenerate at the $K$ point, which is a Dirac point.

The Dirac point can be characterized by the winding number,
defined as an integral along a closed path:
\begin{equation}
  w
  =
  \oint \frac{d\mib{k}}{2\pi}
  \cdot
  \left[ \hat{h}_x^{(\mu)}(\mib{k})\mib{\nabla}\hat{h}_y^{(\mu)}(\mib{k})
    -\hat{h}_y^{(\mu)}(\mib{k})\mib{\nabla}\hat{h}_x^{(\mu)}(\mib{k}) \right],
  \label{eq:winding_number1}
\end{equation}
where
$\hat{\mib{h}}^{(\mu)}(\mib{k})=\mib{h}^{(\mu)}(\mib{k})/|\mib{h}^{(\mu)}(\mib{k})|$
with $\mib{h}^{(\mu)}(\mib{k})=[h^{(\mu)}_x(\mib{k}), h^{(\mu)}_y(\mib{k})]$.
This winding number is independent of the choice of $\mu$,
as the matrix elements of the Hamiltonian for different $\mu$ are related
by the gauge transformation.
We obtain $w=-1$ around $K=2\pi(1/3\sqrt{3},1/3)/a$
and $w=1$ around $K'=2\pi(2/3\sqrt{3},0)/a$.

For a lattice with finite width,
we should choose unit cells that conform to the edge structure
(see Fig.~\ref{graphene_edges}).
For armchair or bearded edges, we should choose $\mu=0$;
for zigzag or twig edges, we should select $\mu=1$ or $2$.
Along the edges, we can define wave number $\mib{k}_{\parallel}$.
It is a one-dimensional vector for the honeycomb lattice:
$\mib{k}_{\parallel}=(k_y)$ for armchair or twig edges and
$\mib{k}_{\parallel}=(k_x)$ for zigzag or bearded edges,
but we will use the same notation for the two-dimensional surface wave number
in the case of the diamond lattice.
The system with a fixed $\mib{k}_{\parallel}$ can be regarded as
a one-dimensional system perpendicular to the edges.
We denote the lattice constant of this one-dimensional system as $\tilde{a}$;
$\tilde{a}=(\sqrt{3}/2)a$ for armchair or twig  edges and
$\tilde{a}=(3/2)a$        for zigzag or bearded edges.
For the existence of edge states in a system with chiral symmetry,
the winding number for this one-dimensional system
with fixed $\mib{k}_{\parallel}$ is important~\cite{Ryu2002, Hatsugai2009}.
It is defined as
\begin{equation}
  \begin{split}
    w(\mib{k}_{\parallel})
    =
    \int_0^{2\pi/\tilde{a}} \frac{dk_{\perp}}{2\pi}
    \biggl[
      &\hat{h}_x^{(\mu)}(\mib{k})
      \frac{\partial}{\partial k_{\perp}} \hat{h}_y^{(\mu)}(\mib{k})\\
      &-\hat{h}_y^{(\mu)}(\mib{k})
      \frac{\partial}{\partial k_{\perp}} \hat{h}_x(\mib{k})
      \biggr],
  \end{split}
  \label{eq:winding_number2}
\end{equation}
where $k_{\perp}$ is the wave number perpendicular to the edges,
i.e.,
$k_{\perp}=k_x$ for armchair or twig edges
and
$k_{\perp}=k_y$ for zigzag or bearded edges.
A nonzero winding number indicates the existence of an edge state.
Note that this quantity is defined in the limit of infinite width,
that is, without edges.
However, we can use this bulk quantity
to discuss the existence of edge states.

First, we consider edges along the $y$ direction.
For the armchair edges ($\mu=0$),
we have $w(\mib{k}_{\parallel})=0$ in the entire region.
For the twig edges,
$w(\mib{k}_{\parallel})=-1$ for $\mu=1$, and
$w(\mib{k}_{\parallel})= 1$ for $\mu=2$ in the entire region.
Next, we consider edges along the $x$ direction.
For the bearded edges ($\mu=0$),
$w(\mib{k}_{\parallel})=1$
for $-2\pi/(3\sqrt{3}a) < k_x < 2\pi/(3\sqrt{3}a)$
and zero otherwise.
For the zigzag edges ($\mu=1$ or $2$),
$w(\mib{k}_{\parallel})=0$
for $-2\pi/(3\sqrt{3}a) < k_x < 2\pi/(3\sqrt{3}a)$
and $-1$ otherwise.

\begin{figure}
  \begin{center}
    \includegraphics[width=0.99\linewidth]
      {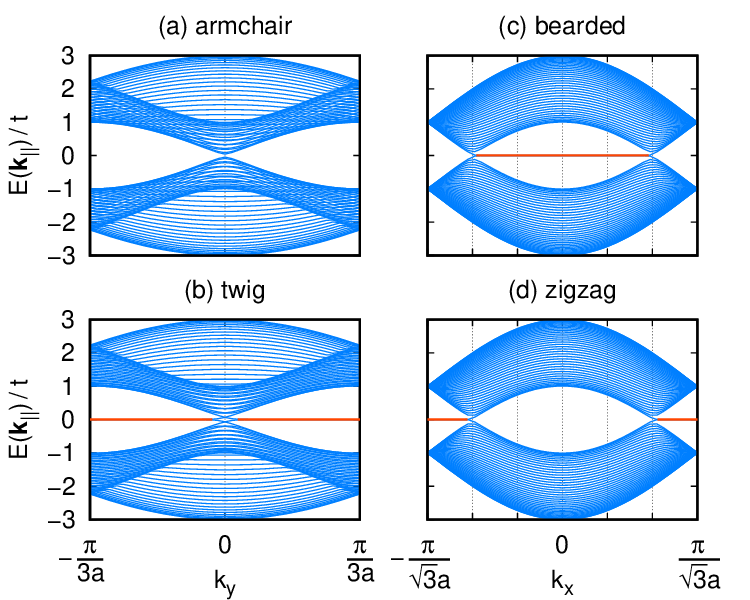}%
  \end{center}
  \caption{
    (Color online)
    Band structures of a honeycomb lattice with finite width
    for (a) armchair, (b) twig, (c) bearded, and (d) zigzag edges.
    The number of lattice sites perpendicular to the edges is 40.
    The zero-energy states in (b)--(d), represented by the red lines,
    are edge states.
    \label{graphene_edge_states}}
\end{figure}

In Fig.~\ref{graphene_edge_states},
we show the band structures of honeycomb lattices with finite width
with different edge shapes.
Except for the armchair edges,
a zero-energy flat band appears.
It is isolated from the bulk band and corresponds to an edge state.
The appearance of the edge state is in accord with the nonzero winding number.

\section{Diamond lattice}
We now discuss the surface states of the tight-binding model
on a diamond lattice with nearest-neighbor hopping.
The primitive translation vectors are
$\mib{a}_1=a(0,1,1)/2$,
$\mib{a}_2=a(1,0,1)/2$, and
$\mib{a}_3=a(1,1,0)/2$,
where $a$ is the lattice constant of the fcc lattice.
The vectors connecting nearest-neighbor sites are
$\mib{\delta}_0=a(1,1,1)/4$,
$\mib{\delta}_1=a(1,-1,-1)/4$,
$\mib{\delta}_2=a(-1,1,-1)/4$, and
$\mib{\delta}_3=a(-1,-1,1)/4$.
We define $\mib{a}^{(\mu)}_i=\mib{\delta}_{\mu}-\mib{\delta}_i$
as in the previous section.
In particular,
$\{ \mib{a}^{(0)}_0, \mib{a}^{(0)}_1, \mib{a}^{(0)}_2, \mib{a}^{(0)}_3 \}
=\{ \mib{a}_0, \mib{a}_1, \mib{a}_2, \mib{a}_3 \}$
with $\mib{a}_0=(0,0,0)$.
Using the above vectors $\mib{a}^{(\mu)}_i$,
the Hamiltonian is written in the same form as Eq.~\eqref{eq:Hamiltonian}
with Eqs.~\eqref{eq:hx} and ~\eqref{eq:hy}.
The choice of unit cells is denoted by $\mu$ as in the previous section.

\begin{figure}
  \begin{center}
    \includegraphics[width=0.8\linewidth]
      {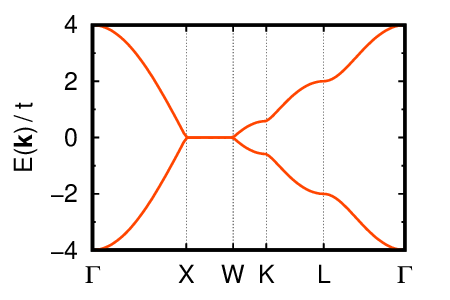}%
  \end{center}
  \caption{
    (Color online)
    Band structure of the diamond lattice model.
    The high symmetry points are
    $\Gamma=(0,0,0)$, $X=2\pi(1,0,0)/a$, $W=2\pi(1,1/2,0)/a$,
    $K=2\pi(3/4,3/4,0)/a$, and $L=2\pi(1/2,1/2,1/2)/a$.
    \label{diamond_lattice_dispersion}}
\end{figure}

The band dispersion is shown in Fig.~\ref{diamond_lattice_dispersion}.
The bands degenerate along the line connecting
$X=2\pi(1,0,0)/a$ and $W=2\pi(1,1/2,0)/a$.
It is a nodal line, which can be characterized
by the winding number Eq.~\eqref{eq:winding_number1}.
For a path encircling this line, the winding number is $w=1$.
For the line $X$--$W'$ [$W'=2\pi(1,0,1/2)/a$],
the winding number is $w=-1$.
In the following subsections,
we discuss the surface states for each surface direction.

\subsection{(001) surface}
\begin{figure}
  \begin{center}
    \includegraphics[width=0.99\linewidth]
      {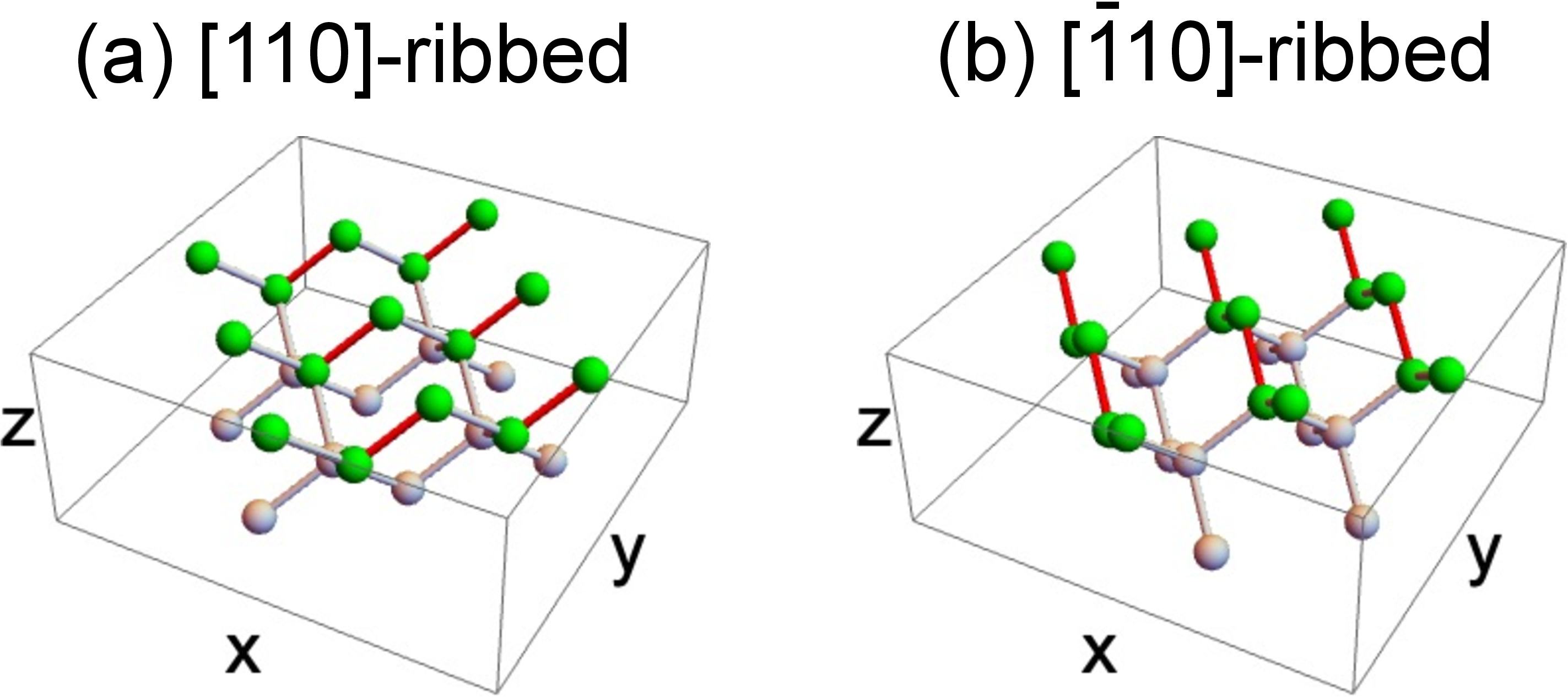}%
  \end{center}
  \caption{
    (Color online)
    Surface structures of the (001) surface of a diamond lattice:
    (a) [110]-ribbed and (b) [$\bar{1}$10]-ribbed.
    Each red bond connects two sites (green spheres)
    that form a unit cell on the surface.
    \label{001_surface}}
\end{figure}

First, we consider surfaces perpendicular to the [001] direction.
There are two types of surface structure, as shown in Fig.~\ref{001_surface}.
The surface shown in Fig.~\ref{001_surface}(a) has a ribbed structure
along the [110] direction,
and thus, we name it the [110]-ribbed surface.
This surface is obtained by choosing unit cell $\mu=0$ or $3$.
Similarly, we name the surface shown in Fig.~\ref{001_surface}(b)
the [$\bar{1}$10]-ribbed surface,
which is obtained by choosing $\mu=1$ or $2$.
These two surfaces are equivalent upon a $90^{\circ}$ rotation
around the $z$-axis.

\begin{figure}
  \begin{center}
    \includegraphics[width=0.99\linewidth]
      {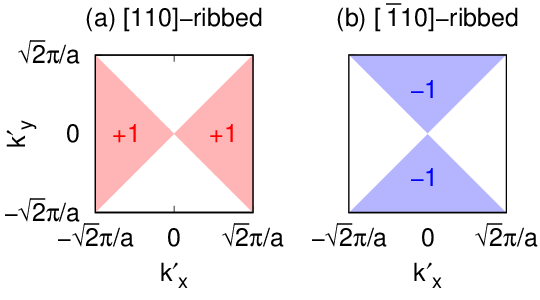}%
  \end{center}
  \caption{
    (Color online)
    Winding number for (001) surfaces
    with (a) [110]-ribbed and (b) [$\bar{1}$10]-ribbed structures.
    The winding number is zero in the unshaded regions.
    \label{winding_number_001}}
\end{figure}

The winding number, as given by Eq.~\eqref{eq:winding_number2}
is calculated for each surface wave number $\mib{k}_{\parallel}$ 
and is shown in Fig.~\ref{winding_number_001}.
Here, $\mib{k}_{\parallel}=(k_x',k_y')$
with $k_x'$ along the [110] direction
and $k_y'$ along [$\bar{1}$10] the direction.
For the calculation of the winding number Eq.~\eqref{eq:winding_number2},
we set $k_{\perp}=k_z$ and $\tilde{a}=a/2$ for the (001) surfaces.
The winding number is nonzero in a part of the Brillouin zone.
The system with the [110]-ribbed surfaces has a nonzero winding number
in the region
where the system with the [$\bar{1}$10]-ribbed surfaces
has a zero winding number, and vice versa.

\begin{figure}
  \begin{center}
    \includegraphics[width=0.8\linewidth]
      {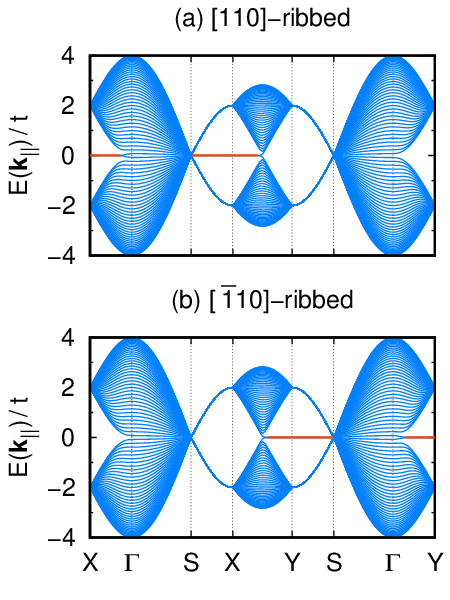}%
  \end{center}
  \caption{
    (Color online)
    Band structures of a diamond lattice with finite thickness
    for (001) surfaces
    with (a) [110]-ribbed and (b) [$\bar{1}$10]-ribbed structures.
    The high symmetry points are
    $\Gamma=(0,0)$, $X=\sqrt{2}\pi(1,0)/a$, $Y=\sqrt{2}\pi(0,1)/a$,
    and $S=\sqrt{2}\pi(1,1)/a$.
    The number of lattice sites perpendicular to the surfaces is 40.
    The zero-energy states, represented by the red lines, are surface states.
    \label{surface_states_001}}
\end{figure}

The band structure for a lattice with finite thickness with (001) surfaces
is shown in Fig.~\ref{surface_states_001}.
The zero-energy surface state appears in the region
where $w(\mib{k}_{\parallel}) \ne 0$.
In Ref.~\citen{Hirashima2016}, based on a discussion of bound states,
a completely flat band was found for BB(001) surfaces defined there.
The BB(001) surfaces are
a combination of [110]-ribbed and [$\bar{1}$10]-ribbed surfaces,
which results in a completely flat band
across the entire surface Brillouin zone. 
The AB(001) surfaces in Ref.~\citen{Hirashima2016}
correspond to the [110]-ribbed surfaces here.
In Ref.~\citen{Hirashima2016},
only a portion of the symmetric lines in the Brillouin zone was investigated,
and the disappearance of the surface state around the $Y$ point
was not discussed.
The existence of the surface state in part of the Brillouin zone
may be difficult to understand using the theory in Ref.~\citen{Hirashima2016},
but it is naturally understood in the present framework by considering
the topological nature of the system.

\subsection{(110) surface}
\begin{figure}
  \begin{center}
    \includegraphics[width=0.99\linewidth]
      {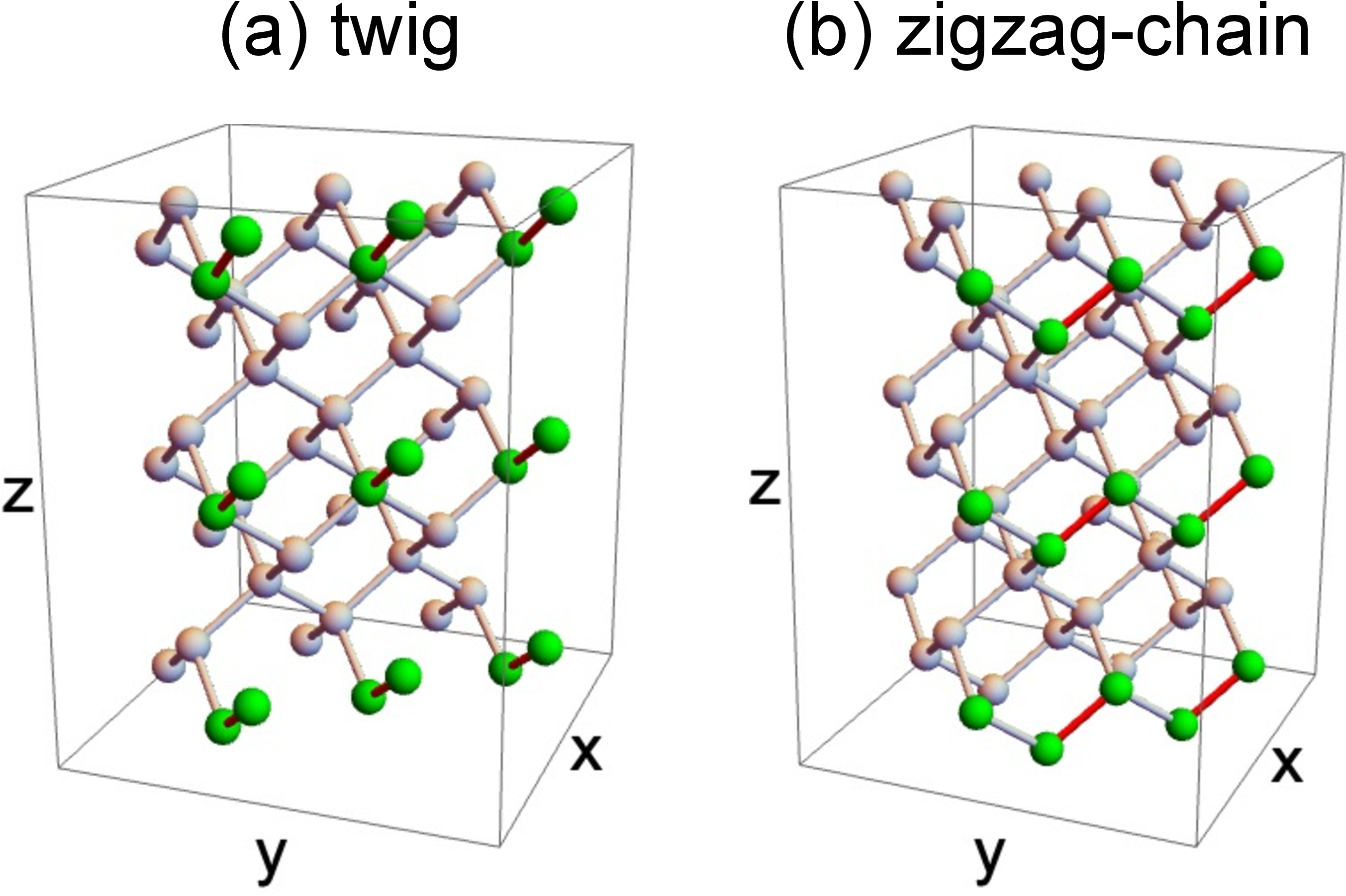}%
  \end{center}
  \caption{
    (Color online)
    Surface structures of the (110) surface of a diamond lattice:
    (a) twig and (b) zigzag-chain.
    Each red bond connects two sites (green spheres)
    that form a unit cell on the surface.
    \label{110_surface}}
\end{figure}

Next, we consider surfaces perpendicular to the [110] direction.
As in the previous case, two types of surface structures are possible,
as shown in Fig.~\ref{110_surface}.
We refer to the surface shown in Fig.~\ref{110_surface}(a) as the twig surface
due to its resemblance to the twig edge of the honeycomb lattice~\cite{Xia2023}.
This surface is realized by choosing the unit cell with $\mu=0$ or $\mu=3$.
We name the surface shown in Fig.~\ref{110_surface}(b) the zigzag-chain surface,
as it consists of zigzag chains running along the [$\bar{1}$10] direction.
This surface is obtained by choosing $\mu=1$ or $\mu=2$.

For the calculation of the winding number Eq.~\eqref{eq:winding_number2},
we use $\mib{k}_{\parallel}=(k_y',k_z)$ with $k_y'$
along the [$\bar{1}$10] direction.
$k_{\perp}$ is along the [110] direction and $\tilde{a}=a/2\sqrt{2}$
for the (110) surfaces.
The winding number $w(\mib{k}_{\parallel})$ for the twig surfaces depends
on the choice of $\mu$:
$w(\mib{k}_{\parallel})=1$ for $\mu=0$ and $w(\mib{k}_{\parallel})=-1$ for $\mu=3$
throughout the entire surface Brillouin zone.
In contrast, for the zigzag-chain surfaces, the winding number is
$w(\mib{k}{\parallel}) = 0$ across the entire surface Brillouin zone.

\begin{figure}
  \begin{center}
    \includegraphics[width=0.8\linewidth]
      {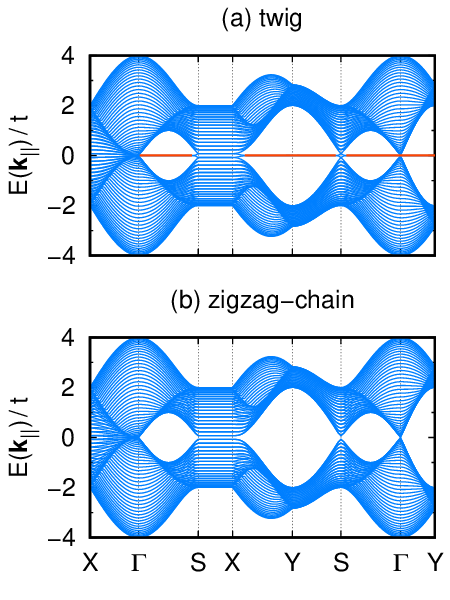}%
  \end{center}
  \caption{
    (Color online)
    Band structures of a diamond lattice with finite thickness,
    terminated by (110) surfaces
    with (a) twig and (b) zigzag-chain structures.
    The high symmetry points are
    $\Gamma=(0,0)$, $X=\pi(\sqrt{2},0)/a$, $Y=\pi(0,1)/a$,
    and $S=\pi(\sqrt{2},1)/a$.
    The number of lattice sites along the direction perpendicular
    to the surfaces is 40.
    The zero energy states represented by the red lines
    in (a) are surface states.
    \label{surface_states_110}}
\end{figure}

Figure~\ref{surface_states_110} shows the band structure of a diamond lattice
with finite thickness, terminated by (110) surfaces.
For the twig surfaces shown in Fig.~\ref{surface_states_110}(a),
zero-energy surface states appear
across the entire Brillouin zone,
consistent with the nonzero winding number,
provided the projected bulk band gap remains open.
In contrast, no surface state is found for the zigzag-chain surfaces,
as shown in Fig.~\ref{surface_states_110}(b),
in agreement with the vanishing winding number.

\subsection{(111) surface}
\begin{figure}
  \begin{center}
    \includegraphics[width=0.99\linewidth]
      {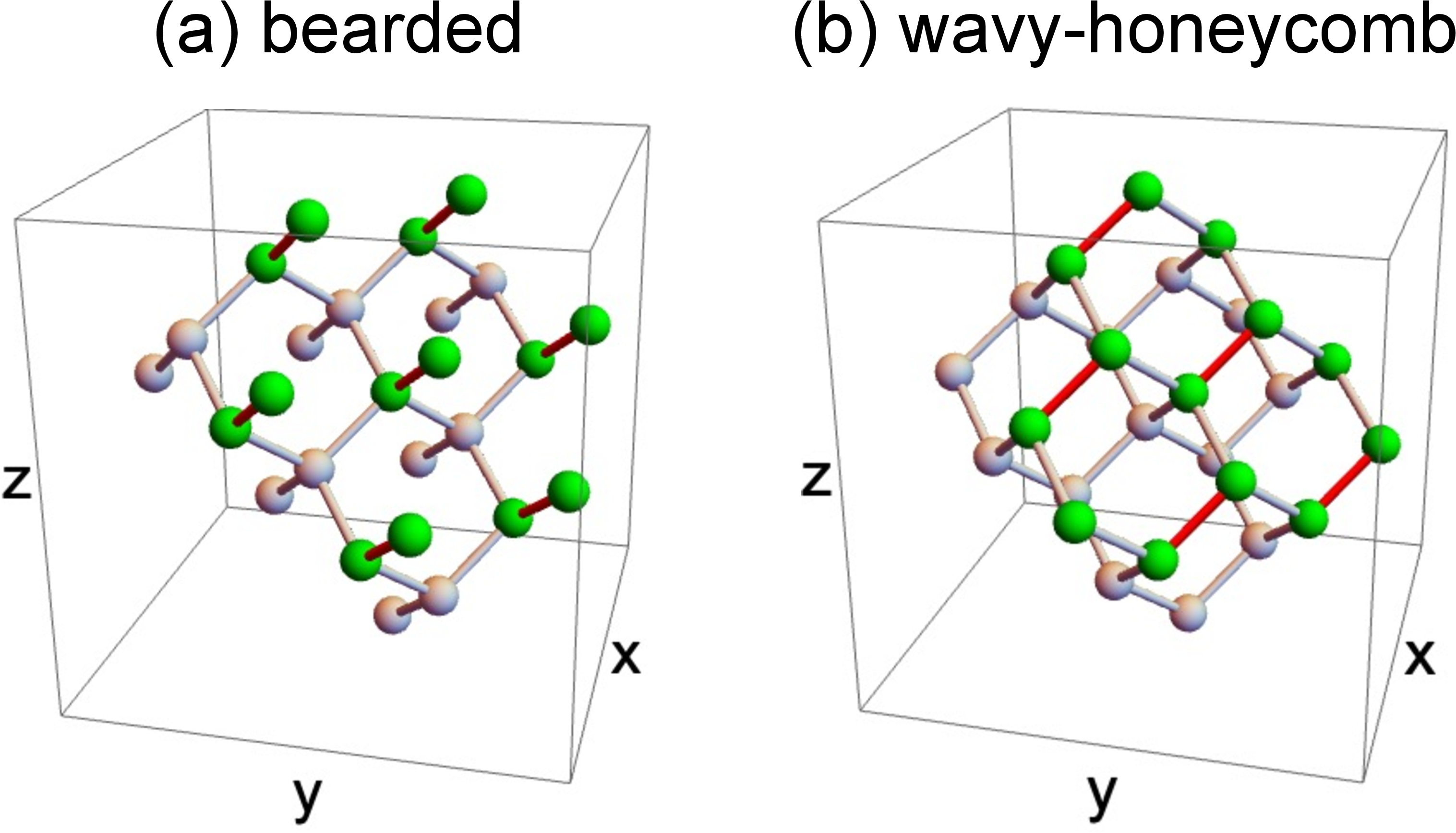}%
  \end{center}
  \caption{
    (Color online)
    Surface structures of the (111) surface of a diamond lattice:
    (a) bearded and (b) wavy-honeycomb.
    Each red bond connects two sites (green spheres)
    that form a unit cell on the surface.
    \label{111_surface}}
\end{figure}

Finally, we consider surfaces perpendicular to the [111] direction.
As in the previous cases,
two types of surface structures are possible,
as shown in Fig.~\ref{111_surface}.
We refer to the surface shown in Fig.~\ref{111_surface}(a)
as the bearded surface
due to its structural similarity to the bearded edge
of the honeycomb lattice~\cite{Ryu2002}.
This surface is realized by choosing the unit cell with $\mu=0$.
We name the surface shown in Fig.~\ref{111_surface}(b)
the wavy-honeycomb surface,
which reflects its undulating honeycomb structure.
This surface is obtained by choosing $\mu=1$, $2$, or $3$.

\begin{figure}
  \begin{center}
    \includegraphics[width=0.99\linewidth]
      {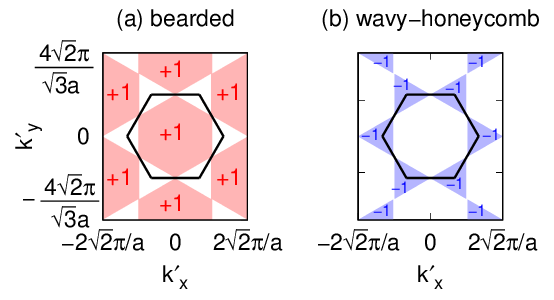}%
  \end{center}
  \caption{
    (Color online)
    Winding number for the (111) surfaces
    with (a) bearded and (b) wavy-honeycomb structures.
    The winding number is zero in the unshaded regions.
    The region surrounded by the bold line is the first Brillouin zone.
    \label{winding_number_111}}
\end{figure}

The winding number $w(\mib{k}_{\parallel})$ is shown
in Fig.~\ref{winding_number_111},
where $\mib{k}_{\parallel}=(k_x',k_y')$ with
$k_x'$ is along the [$\bar{1}$10] direction
and
$k_y'$ is along the [$\bar{1}\bar{1}$2] direction.
$k_{\perp}$ is along the [111] direction and $\tilde{a}=a/\sqrt{3}$
for the (111) surfaces.
For the bearded surfaces,
the winding number is nonzero in regions
where the winding number for the wavy-honeycomb surfaces vanishes,
and vice versa.

\begin{figure}
  \begin{center}
    \includegraphics[width=0.8\linewidth]
      {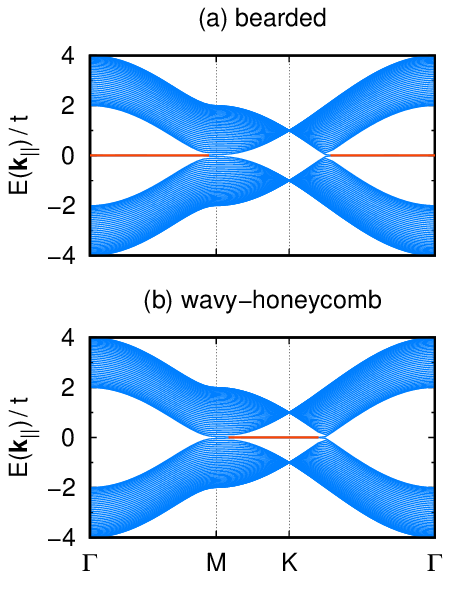}%
  \end{center}
  \caption{
    (Color online)
    Band structures of a diamond lattice with finite thickness
    for (111) surfaces
    with (a) bearded and (b) wavy-honeycomb structures.
    The high symmetry points are
    $\Gamma=(0,0)$, $M=2\sqrt{2}\pi(0,1/\sqrt{3})/a$,
    and $K=2\sqrt{2}\pi(1/3,1/\sqrt{3})/a$.
    The number of lattice sites along the direction
    perpendicular to the surfaces is 40.
    The zero energy states represented by the red lines are the surface states.
    \label{surface_states_111}}
\end{figure}

Figure~\ref{surface_states_111} shows the band structure
of a diamond lattice with finite thickness, terminated by (111) surfaces.
The regions where the surface states appear are consistent with the condition
$w(\mib{k}_{\parallel}) \ne 0$.
The (111) surfaces discussed in Ref.~\citen{Takagi2008}
and the AB(111) surfaces studied in Ref.~\citen{Hirashima2016}
correspond to the wavy-honeycomb surfaces described here.
The surface states for the wavy-honeycomb surfaces
are consistent with these previous studies.
A topological explanation for the presence of the surface states
for the wavy-honeycomb surfaces
can also be found in Ref.~\citen{Takahashi2013}.
Additionally, Ref.~\citen{Hirashima2016} reported a completely flat band
for BB(111) surfaces.
These surfaces are a combination of bearded and wavy-honeycomb surfaces,
resulting in a completely flat band across the entire surface Brillouin zone.

\section{Summary}
We have investigated the surface states
of a tight-binding model with nearest-neighbor hopping
on a diamond lattice,
with particular focus on the influence of surface structure
for all the (001), (110), and (111) surfaces.

In lattices with two sublattices, such as the diamond lattice,
the surface structure is not uniquely determined
even when the surface direction is fixed
because we can choose which sublattice is terminated on the surface.
Our results show that the existence of surface states depends
not only on the direction of the surface,
but also crucially on its structure.
For each surface direction,
we identified two distinct types of surface structures.
We find that the regions of the surface Brillouin zone
where surface states appear differ between these two surface structures.
In particular, for the (110) surface,
no surface states appear for the zigzag-chain surface,
while for the twig surface,
a flat surface state spans the entire surface Brillouin zone.
The presence of such flat surface states leads
to an enhanced density of states near the surface,
which can trigger electronic instabilities
such as magnetism or superconductivity
when electron-electron interactions are taken into account.
Indeed, in a two-dimensional case, edge magnetism has been discussed
for the honeycomb lattice~\cite{Fujita1996}.
A possible enhancement of the superconducting transition temperature
due to flat surface states has also been proposed in Ref.~\citen{Kopnin2011}.

We also note that
the existence of surface states is determined
by the winding number $w(\mib{k}_{\parallel})$,
which is calculated from the matrix elements of the bulk Hamiltonian.
In this sense, the presence or absence of surface states is governed
by the bulk Hamiltonian.
However, the matrix elements depend on the choice of the unit cell,
which should be chosen to match the surface structure.
Thus, in this sense, the existence of surface states
depends on the surface structure.

\begin{acknowledgments}
The author thanks D. S. Hirashima for fruitful discussions.
This work was supported by JSPS KAKENHI Grant No.
JP23K03330. 
\end{acknowledgments}


\end{document}